\documentstyle[aps,prl,amssymb,psfig,12pt]{revtex}
\begin{document}
%\draft % makes pacs numbers print
\title{Lyapunov Exponents and
KS Entropy for the Lorentz Gas
at Low Densities}
\author{Henk van Beijeren}
\address{Institute for Theoretical Physics, Utrecht University,
P.O. Box 80006, 3508 TA Utrecht, The Netherlands}
\author{J. R. Dorfman}
\address{IPST, and Department of Physics,University of Maryland, College Park,
MD 20742}
\date{\today}
\maketitle
\begin{abstract}
The Lyapunov exponents and the Kolmogorov-Sinai entropy for a
two dimensional Lorentz gas at low densities are defined for general
non-equilibrium states and calculated with the use of
a Lorentz-Boltzmann type equation. In equilibrium the density dependence of
these
quantities predicted by Krylov, is recovered and  explicit
expressions are obtained . The
relationship between KS entropy, Lyapunov exponents and escape rate,
conjectured by Gaspard and Nicolis for non-equilibrium systems, is confirmed
and generalized.
\end{abstract}
%\narrowtext

  A standard model for studying irreversible processes in classical fluids
is the Lorentz
 gas, consisting of a system of fixed hard sphere ( $ d = 3 $)
or hard disk ($ d=2 $) scatterers placed at random in space, with a particle
that moves freely between elastic collisions with the scatterers{\cite
{hauge}}. This
simple model of diffusion has been the object of a great deal of interest,
as a non trivial system with irreversible behavior, that is accessible both
to careful mathematical analysis,
and to a study of its transport
properties
by means of kinetic theory.
It has been possible to provide
rigorous proofs that,
under reasonable physical
assumptions, the
Lorentz gas is at least a K- system and that the periodic
Lorentz gas is a Bernoulli system \cite{sinai1,buni1}.
This implies that the Lorentz gas has
a well defined equilibrium state and that a suitably
defined initial distribution will approach equilibrium.

   The purpose of this letter is to present a calculation of quantities that
characterize the dynamic properties of  random
 Lorentz
gases. We illustrate the method for the two dimensional case
and compute
the positive Lyapunov exponent
$ \lambda^+ $
and the Kolmogorov - Sinai
entropy, $ h_{KS} $, for such a model
in the limit of low density of scatterers
for two cases: (1) The system is large and has periodic boundaries ( which we
eventually allow to move off to infinity); and (2) The scatterers are
distributed randomly over a
large finite area with an absorbing boundary.
The first case will allow us to verify a conjecture of Krylov,
 discussed at some length by Sinai and others~\cite{krylov},
that $ \lambda^+
\sim \tilde{n}
\ln \, \tilde{n}
 $ , where $ \tilde{n} $ is the reduced density
of scatterers, $ \tilde{n} = na^2 $,
with $ n $ the density of
the scatterers, and $ a $ their radius. The second case will allow us
to to verify a conjecture of Gaspard and Nicolis~\cite{gasni,gaspbaras} that
the Lyapunov exponents
and Kolmogorov - Sinai entropy for open systems have finite size corrections
that can be related to the coefficient of diffusion.

  The starting point for our analysis is a result due to Sinai~\cite{sinai1},
for the
curvature of an expanding "wave front" that describes the unstable
manifold of the phase space trajectory for the moving particle in the
positive time direction.
For our purposes the essential ingredients in Sinai's formula
are: (a) The curvature, $ \kappa $, is the inverse of
the radius of curvature, $ \rho $ , i.e., $ \kappa = ( \rho )^{-1} $.
In the present, two dimensional case $\rho$ may be interpreted simply as
the distance of two infinitesimally close particle trajectories to their
mutual intersection point, which, for the unstable manifold, has to be
located in the backward direction.
(b) Between collisions $ \rho $ increases linearly with time as $ vt $
where $ v $ is the (constant) speed of the moving particle.
(c) Whenever
the particle collides with a scatterer, there is an instantaneous change
in the curvature according to $ \kappa^{+} = \kappa^{-} + \frac{2}{a cos\phi}$
 where $ \kappa^{+}
=1/\rho^+ $ is the curvature immediately after the collision,
 $ \kappa^{-}
=1/\rho^-$ is the curvature immediately before the collision, and
 $ \phi $ is the angle of incidence in the collision of the particle
 with a scatterer, with $ -{\pi}/{2} \leq \phi \leq {\pi}/{2} $
(see Fig. 1).
\begin{figure}[b]
\centerline{\psfig{figure=lorentz.eps,angle=-90,height=5cm}}
\caption{The change in the radius of curvature at a collision}
\end{figure}
Suppose the trajectories go through collisions at times
$s_1,s_2,\cdots,s_n,\cdots$. Then the radius of curvature at time $t$ between
$s_n$ and $s_{n+1}$ satisfies the relation
\begin{equation}
 \rho(t) = v\tau + \left(\frac{2}{acos\phi_n} +\frac{1}{
\rho^{-}_n}\right)^{-1}
\label{recur}
\end{equation}
where $ \tau = t - \sum_{1}^{n}\tau_j  $,
 and $ \rho^{-}_n $ is the radius
of curvature immediately before the collision with scatterer $ s_n $.

  The relation between the curvature $ \rho(t) $ and the positive Lyapunov
exponent
 characterizing a particular trajectory of the moving particle  for the
two dimensional case discussed here, is given by the time average
\begin{equation}
 \lambda^+({\bf r},{\bf v}) = \lim_{t \rightarrow \infty} \frac{v}{t}
   \int_0^t \frac{1}{\rho({{\bf r}(\tau),{\bf v}(\tau)},\rho_0,\tau)}d\tau
\end{equation}

In eq.(2),
the initial radius of curvature
$\rho_0$ may be taken to be any convenient
value, since the initial value will be unimportant for the time average. In
general this time average
will be difficult to compute. However if the system is ergodic, the time
average can be replaced by an
ensemble average over all allowed positions and velocities of the
moving particle, i.e.
\begin{equation}
   \lambda^+  =v \langle  \frac{1}{\rho({\bf r,v})}  \rangle
\label{lambda}
\end{equation}
where $ \rho({\bf r,v}) $
is the limit of $\rho({\bf r,v},t)$ for fixed final
coordinates and $t \rightarrow \infty$ (this limit is
independent of $\rho_0$). A further
important simplification is obtained by averaging also over all allowed
positions of scatterers (which we will assume to be non-overlapping). In that
case we obtain an expression for $\lambda^+$ of the same form as
eq.(\ref{lambda}), but now the brackets imply an average over the scatterer
positions as well as over the position and velocity of the moving particle.
It is this expression that we wish to evaluate for the cases described above.
%when
For the case of a
large system with periodic boundaries, there is no conceptual difficulty with
imagining
$ \rho $ to be generated by an
infinite number of collisions in the past. For the open system, the
construction of Gaspard and Nicolis
requires that we consider
the set of trajectories on
the fractal repeller,
that is, the set of trajectories
that remain forever trapped within the confines of the
absorbing boundary.
On this set of trajectories
we indeed can imagine an infinite number of
past collisions
generating $ \rho $.

   We calculate $ \lambda^+ $ using eq.(\ref{lambda}) by expressing the average
value in terms of a
distribution function $ P({ \bf r}, {\bf v}, \rho,t) $ such that
$ Pd{\bf r}d{\bf v}d \rho $
is the probablility of finding a particle in the indicated ranges of variables
at time $ t $ . The Lyapunov
exponent can be expressed in terms of $ P $ as
\begin{equation}
\lambda^+ = \frac{v}{ N(t)}\int \frac{1}{\rho}P({\bf r},{\bf v},\rho,t)\,d{\bf
r} \, d{\bf v} \, d\rho
\label{lamkin}
\end{equation}
where
$ N(t) = \int P \, d{\bf r} \, d{\bf v} \, d\rho$.
For the closed system the function $ P $ does not depend on $ {\bf r}, {\bf
v},$
or $ t $, but
for the open system it does,
as we shall see presently.

The probability distribution $P$ changes in time both through free streaming of
the moving particle
and through its collisions with the scatterers. In a low density Lorentz gas
these collisions occur
with an average frequency $\nu = 2anv$.
A collision of the moving particle
with
a scatterer
produces an instantaneous change in the radius of curvature according to the
result quoted in point
(c) above. As a result of these considerations one can easily show that the
time evolution of the
probability distribution $P$ is described by a Lorentz-Boltzmann type of
equation,  of the form
\begin{eqnarray}
\{\partial/\partial t + {\bf v}\cdot\nabla
\}P({\bf r},{\bf v},\rho,t)
= -v\partial/\partial\rho P({\bf r},{\bf v},\rho,t)+ \nonumber\\
+ \nu\{ -P({\bf r},{\bf v},\rho,t)
+1/2\int_{-\pi/2}^{\pi/2}d\phi\int_0^{\infty}d\rho'\cos\phi \nonumber\\
\delta(\rho - \frac{a/(2\cos\phi)}
{1 + a/(2\rho'\cos\phi)})P(\bf r,\bf v',\rho',t) \}
\label{kin}
\end{eqnarray}
The first term on the right hand side of eq.(\ref{kin}) describes the change in
$\rho$ due to free streaming
and the last two terms describe the changes in $P$ resulting from collisions of
the moving particle
with scatterers. As usual there is a loss term, which in the present case
assumes the simple form
$ -\nu P$ because both the speed of the moving particle and the density of the
scatterers are
constants. The gain term counts collisions transforming precollisional
coordinates $(\bf r',\bf v',\rho')$ into postcollisional coordinates
$(\bf r,\bf v,\rho)$. The relationship between $\bf v$ and $\bf v'$ is given
by ${\bf v}' = {\bf v} -2({\bf v}\cdot\hat{\bf \phi})\hat{\bf \phi} $
, where
$\hat{\bf \phi}$ is the unit vector in the direction from the center of the
scatterer to the point of impact of the moving particle at the collision (see
Fig.1). The usual Lorentz-Boltzmann equation~\cite{hauge} is obtained from
eq.(\ref{kin}) by
integrating
it over all
values of $\rho$, provided $P$ satisfies the condition that $P \rightarrow 0$
as
$\rho \rightarrow 0$. We will require that the solutions to eq.(\ref{kin})
satisfy this
condition, since the free streaming current always leads to larger $\rho$, and
there is no influx at the origin from negative values of $\rho$.

  In this paper we only want to solve eq.(\ref{kin}) to lowest order in a
systematic
expansion in powers of the reduced density of scatterers, $\tilde{n} = na^2$.
Since the typical values of $\rho'$ in eq.(\ref{kin}) are of the order of the
mean
free path
$\ell = v/\nu = 1/(2na)$,
the term $a\cos\phi/\rho'$ occurring in the
denominator in the delta function in eq.(\ref{kin}) is of order $\tilde{n}$ and
may be
neglected compared to $1$. The corrections can be shown to be of
relative order
$\tilde{n} \,
\ln \, \tilde{n}$. Then the integration over $\phi$ can easily be
performed and eq.(\ref{kin}) reduces to the somewhat
simpler form
\begin{eqnarray}
&&\{\partial/\partial t + {\bf v}\cdot\nabla\}P({\bf r},{\bf v},\rho,t)=
\nonumber\\
&&= -v\partial/\partial \rho \, P + \nu\{ -P +\frac{1}{a}\Theta(1-\sigma)
\frac{\sigma}{(1 - \sigma^2)^{1/2}}\nonumber\\
&&\:\:\:\:\int_0^{\infty} d\rho'[P(\rho',
{\bf r},{\bf v}_{+}',t) + P(\rho',{\bf r},{\bf v}_{-}',t)]\}
\label{lb}
\end{eqnarray}
where $\Theta$ is the unit step function and $\sigma=2\rho/a$.
The velocities ${\bf v}_{\pm}'$ both
are precollisional velocities, with scattering vectors $\hat{\bf\phi}_{\pm}$
satisfying ${\bf v}\cdot\hat{\bf \phi}_{\pm}=\sigma$.

Let us first consider the equilibrium system
with periodic boundary conditions. In this case the solution to eq.(\ref{lb})
does
not depend on $\bf r,\bf v$ or $t$, and this equation reduces to
\begin{equation}
(v\partial/\partial \rho + \nu)P(\rho) =
\frac{2 \,\nu \,\sigma\Theta(1-\sigma)}{a(1 -
\sigma^2)^{1/2}}\int_0^\infty d\rho'P(\rho')
\end{equation}

The solution to
this equation is simply obtained as $P(\rho) = c_{0}n^e_{m}f_{0
}(\rho)
\delta(|{\bf v}|-v)$ where
$n^e_m$ is the equilibrium density of the moving particle, and  $c_0 =
1/(2\pi v)$ is the normalization of the velocity distribution. To lowest
order in $\tilde{n}$, $f_0$ is
\begin{eqnarray}
f_{0}(\rho) &=& (1/\ell)e^{-\rho/\ell} \qquad\:\:\:\:\:\:\qquad \qquad  \rho >
a/2\nonumber \\
f_{0}(\rho) &=& (1/\ell)[ 1 - (1 - \sigma^2)^{1/2}] \qquad \:\:\:  \rho < a/2
\label{f0}
\end{eqnarray}

Notice that up to corrections of relative order $\tilde{n}$, this solution is
continuous at $\rho = a/2$,
and satisfies the proper normalization condition,
$\int_{0}^{\infty}P(\rho)d\rho = c_{0}n^e_m
\delta(|{\bf v}|-v)$.

 From eqs.(\ref{lamkin})
and (\ref{f0}) the Lyapunov exponent follows immediately as
\begin{equation}
\lambda^+ = 2nav(1 - \ln 2 - C - \ln\tilde{n})      \qquad $for$\; \tilde{n}
\ll 1
\label{krylov}
\end{equation}
where $C$ is Euler's constant. Since for a closed system the Kolmogorov-Sinai
entropy is equal to the
sum of the positive Lyapunov exponents, and since the two dimensional Lorentz
gas has only one
positive Lyapunov exponent, $h_{KS}$ is equal to $\lambda^+$ as given by
eq.(\ref{krylov}). This result is,
of course, in agreement with the conjecture of Krylov, and we have determined
both the coefficient of
the $\tilde{n} \, \ln \, \tilde{n}$ term
and the first correction
to it. The form as well as the coefficient of the $
\tilde{n}\, \ln \,\tilde{n}$
term agrees with the result obtained by Friedman, Oono, and
Kubo \cite{fok} and by Chernov \cite{chernov1} for the KS entropy of
the periodic Lorentz gas as the radii of the scatterers become small.

  Next we turn to the calculation of the Lyapunov exponent for the fractal set
of trapped trajectories
on a finite area with absorbing boundary. Since almost every trajectory of the
moving particle starting
on this area
leads to escape, we
have to construct a non-equilibrium
solution of eq.(\ref{lb}) and compute $\lambda^+$ using eq.(\ref{lamkin}),
taking the limit $t \rightarrow \infty$. The
appropriate solution of eq.(\ref{lb}) is the Chapman-Enskog hydrodynamic
solution~\cite{cc} which is completely determined
by the local density of the moving particle, and the density gradients. Thus,
we look for solutions
of the form
\begin{eqnarray}
P({\bf r},{\bf v},\rho,t)&=&c_0 \delta(|{\bf v}|-v)\{ p_{0}({\bf r},{\bf
v},\rho,t) + \nonumber\\
&&+p_{1}({\bf r},{\bf v},\rho,t) + p_{2}({\bf r},{\bf v},\rho,t)+\cdots\}
\end{eqnarray}
where $p_{i}$ is proportional to the $i-th$ gradient of the local density of
the moving particle. The lowest
order solution is the local equilibrium solution
\begin{equation}
p_{0}({\bf r},{\bf v},\rho,t) =n_m({\bf r},t) f_{0}(\rho)
\label{p0}
\end{equation}
where $f_{0}$ is given by
eq.(\ref{f0}). The first order equation is obtained by consistently keeping all
terms
in eq.(\ref{lb}) which are first order in the gradients, and is given by
\begin{eqnarray}
&&f_{0}(\rho) \, (\partial^{(1)}/\partial t + {\bf v}\cdot\nabla) \, n_m({\bf
r},t) = \nonumber\\
&&= -v\partial/\partial \rho \, p_{1}({\bf r}, {\bf v},\rho,t) + \nu[ -
p_{1}({\bf r},{\bf v},\rho,t) +
\frac{1}{a}\Theta (1-\sigma)\nonumber\\
&& \:\:\:\frac{\sigma}{(1 - \sigma^2)^{1/2}}
\int_{0}^{\infty} d\rho'[p_{1}({\bf r},{\bf v}_{+}',\rho',t) + p_{1}({\bf
r},{\bf v}_{-}',\rho',t)]
\end{eqnarray}
Here the term $\partial^{(1)}/\partial t \, n_m(\bf r,t)$ is the first order
gradient term in the hydrodynamic
equation for the local density of the moving particle. This term is zero, and
is dropped. As the collision
operator does not change the tensorial character in $\bf v$ of the functions on
which it operates, we may
set $p_{1}({\bf r},{\bf v},\rho,t) = f_{1}(\rho) \, {\bf v}\cdot\nabla n_m({\bf
r},t)$, with $f_{1}$ satisfying
\begin{eqnarray}
&&f_{0} + (v\partial/\partial \rho +\nu)f_{1} = \Theta
(1-\sigma)\frac {2\nu\sigma(1 - 2\sigma^2)}
{a(1 - \sigma^2)^{1/2}}\int_0^{\infty} d\rho'f_{1}(\rho')  \nonumber\\
&&
\label{fir}
\end{eqnarray}

The solution to eq.(\ref{fir}) can be determined to lowest order in $\tilde{n}$
by noticing that $f_1$ must
be continuous at $\rho = a/2$ since the right hand side of the equation
contains no delta functions. We then
find that
\begin{eqnarray}
f_{1}(\rho) &=& -(1/v\ell)\rho e^{-\rho/\ell} + (1/4v)e^{-\rho/\ell}
\qquad\qquad \rho > a/2\\
f_{1}(\rho) &=& (1/4v)\{1 - (1-\sigma^2)^{1/2}(1 +2\sigma^2)\} \:\: \qquad \rho
< a/2
\end{eqnarray}
Again, we have used the condition that $f_{1}(\rho) \rightarrow 0$ as $\rho
\rightarrow 0$.
  Proceeding to second order, we find that application of
${\bf v}\cdot\nabla$ to $p_1$ leads
to a term that as a function of $\bf v$ can be separated into a traceless
tensor of the second degree
in $\bf v$ and a scalar part. Consequently $p_2$ can be separated into a
scalar part $p_{2}^{s}$ and
a part proportional to the traceless tensor ${\bf v}{\bf v} - (v^{2}/2){\bf
1}$. For the determination
of the Lyapunov exponent only the scalar term is important since the traceless
tensor part yields
zero on integration over $\bf v$. Thus we write
\begin{equation}
p_{2}^{s}({\bf r},{\bf v},\rho,t) = {\nabla}^{2}n_m({\bf r},t)f_{2}(\rho)
\label{p2}
\end{equation}
and we find that $f_{2}(\rho)$ satisfies the equation
\begin{equation}
f_{0}(\rho)D + (v^{2}/2)f_{1}(\rho) + (\nu + v\partial/\partial
\rho)f_{2}(\rho) = 0
\end{equation}
Here we have imposed a solubility condition in the Chapman-Enskog method which
requires that
$\int_{0}^{\infty}f_{2} \, d \rho = 0$. The quantity $D$, appearing in this
equation is the low density value
of the coefficient of diffusion of the moving particle, which is given by
$D= (3/8)\ell v$.

 From eqs.(\ref{lambda},\ref{p0} and \ref{p2}) one finds the positive
Lyapunov exponent as
\begin{equation}
\lambda^+=\lambda^+_0 +\
\lim_{t\rightarrow \infty}\frac{\kappa_2 \, v \int \nabla^2 n_m({\bf r},t)
d\,{\bf r}}
{N(t)}
\label{GN}
\end{equation}
where ${\lambda}_0$ is the closed system value, (\ref{krylov}),
and\\ $\kappa_2=\int_0^\infty 1/\rho \, f_2(\rho) \, d\, \rho = -\ell/4$.\\
Using Fick's law expressing the diffusion current as ${\bf j}({\bf r},t) =
-D\nabla n_m({\bf r},t)$, defining the escape rate as $E=\lim_{\,t\rightarrow
\infty}
1/N(t)\int_\Lambda \bf
j({\bf r},t) \cdot {\bf \hat{n}} \, dS$, with $\Lambda $ denoting the boundary
of the
system and ${\bf \hat{n}}$ the unit vector pointing outward from the boundary,
and employing Gauss's law, one can reexpress $\lambda^+$ in terms of
$E$ as
\begin{equation}
\lambda^+= \lambda^+_0 -\kappa_2 \, v  E/D =\lambda^+_0+(2/3) \, E.
\label{lamne}
\end{equation}

 We have thus shown that the Lyapunov exponent for a random, two dimensional
Lorentz gas can indeed
  be calculated for both closed and open systems, that this calculation can be
done using standard
  methods from the kinetic theory of gases, and that the results are consistent
with the escape-rate
formalism employed by Gaspard and Nicolis in their discussion of
diffusion \cite{gasni}.
Finally we can obtain the KS entropy for the system with absorbing boundaries
from the well-known relation \cite{ruek} $h_{KS}=\sum_\alpha '
\lambda_\alpha -E$, where the sum runs over the positive Lyapunov exponents
only. Collecting the previous results we find
\begin{equation}
h_{KS}=h^{(0)}_{KS} -(1/3) \, E
\end{equation}
with $h^{(0)}_{KS}$ the KS entropy of the closed system.

 We conclude with a few brief remarks. (1) This method can be extended in a
number of directions -
to higher dimensions, to higher densities, and to more complicated processes
where all of the
particles are moving. One needs to work out the kinetic theory for the
appropriate Boltzmann-like
equation to do so. Work in these directions is in progress.
(2) One of the most remarkable conclusions
to be drawn is that for a
two dimensional Lorentz gas the Lyapunov exponents and KS entropy can be
expressed as ensemble averages of a static local quantity,
the local
curvature $1/\rho({\bf r},{\bf v})$. In how far this can be generalized to
higher dimensional systems is presently under investigation. For the two
dimensional Lorentz gas
Lyapunov exponents and KS entropy
can be defined for arbitrary non-equilibrium ensembles through
eq.(\ref{lambda}), and the approach to equilibrium of these quantities can
be obtained from the time evolution of these ensembles.
(3) In our derivation of the relationship (\ref{lamne}) between the
nonequilibrium positive Lyapunov exponent and the escape rate we only used
Fick's law and nowhere did we have to specify the precise nature of the
boundary conditions.
Hence, this relationship is generally
valid, as long as the state of the system can be described by a
Chapman-Enskog type hydrodynamic distribution function.
The system size must be large, however, so that higher order gradient
and boundary layer corrections may be neglected. We may also conclude that
deviations from the
equilibrium values occur only
for a non-vanishing escape rate, at
least in the absence of external force fields.
(4) Our results
are the
analogs for a continuous system of closely related calculations for Lorentz
lattice gases by
Ernst, Dorfman, Nix and Jacobs, reported in a companion paper~\cite{edjn}.
The Lorentz lattice gas is considerably
simpler to study since it can be studied as a Markov chain, and it is very
amenable to computer
simulations. The two systems are very close in spirit, and the methods to treat
them have many
similar features. Finally, it is a pleasure to note that Boltzmann's equation
is so helpful in
determining some of the main features of the chaotic dynamics of many particle
systems that are
ultimately responsible for the irreversible behavior that Boltzmann
understood at a very deep level.

 The authors would like to thank E.G.D. Cohen, M.H. Ernst, P.
Gaspard, C. Haskell, K. Ide, T. Kirkpatrick, A. Latz, and D. Thirumalai for
helpful and stimulating remarks during various stages of this work.
J.R.D. acknowledges support from the National Science Foundation under
Grant PHY 93-21312. H. v.B. thanks the IPST of the University of Maryland at
College Park for its hopitality and support.

\end{document}